%% file: synchronized_fiber_network.tex
\documentclass[a4paper,11pt]{article}

\usepackage{jinstpub}

\usepackage[noabbrev]{cleveref}
\usepackage{subcaption}
\usepackage{acronym}
\usepackage{bytefield}
\usepackage{circledsteps}
\usepackage[binary-units=true]{siunitx}
\usepackage{tikz}
\usetikzlibrary{calc,decorations.markings,patterns,arrows}
\usepackage{tikz-timing}
\usetikztiminglibrary{advnodes,clockarrows}

\DeclareSIUnit\logicelements{\acsp{le}}

\title{FPGA-based low-cost synchronized fiber network for experimental setups in space}

\author[a]{Tim Oberschulte}
\author[b]{Thijs Wendrich}
\author[a]{Holger Blume}

\affiliation[a]{Institute of Microelectronic Systems, Leibniz Universität Hannover,\\Appelstraße 4, 30167 Hannover, Germany}
\affiliation[b]{Institute of Quantum Optics, Leibniz Universität Hannover,\\Welfengarten 1, 30167 Hannover, Germany}
\emailAdd{tim.oberschulte@ims.uni-hannover.de}

\abstract{Custom experiment setups in physics often require control electronics to execute actions and measurements on a small time scale.
    When further constraints limit the experiment's environment, for example when the experiment is inside a sounding rocket, conventional network systems will not suffice those constraints because of weight, heat or budget limitations.
    This paper proposes a network architecture with a time resolution of less than \SI{1}{\nano\second} over a pair of plastic fibers while using low-cost commercial hardware.
    The plastic fibers in comparison to copper fibers have a low weight and additionally can isolate parts of the setup galvanically.
    Data rates of \SI{40}{\mega\bit\per\second} enable the network to transfer large amounts of measurements and configuration data over the network.
    Proof-of-concept implementations of network endpoints and switches on small FPGAs are analyzed in terms of synchronicity, data rate and resource usage. Using commercial parts the resolution of \SI{1}{\nano\second} is reached with a standard deviation of less than \SI{100}{\pico\second}. Compared to a copper wire implementation the weight is reduced by about one order of magnitude.
    With its low weight at a low cost, the network is useful in space or laboratory setups which require high time resolution.}


\begin{document}

\maketitle

\section{Introduction}
    Complex control electronics are often required for modern physics experiments. In bigger setups, those electronic components may even be located far away from each other but need to communicate and execute commands with precise isochronous timing.
    When experimenting in limiting surroundings, like aboard a sounding rocket or in orbital flight in space, additional constraints will come along, e.g. weight limits or system heat dissipation limitations. Two missions where such limitations apply are discussed here.

    In 2017 the first \ac{bec} has been created on board a sounding rocket in MAIUS-A \cite{becker2018space}. In two follow-up missions, the MAIUS-B \cite{grosse2017maius} and \ac{beccal} \cite{frye2021bose} projects, two different atomic species will be used for creating BECs aboard a rocket and on the \ac{iss}. For those two new projects several new requirements arose in terms of weight, routing complexity, isolation, synchronicity, data rates and financial budget.

    As an additional laser system and other new parts will increase the volume and mass of the design, several parts have to become smaller and lighter to fit into the rocket and not exceed its take-off weight. The approach to improve the communication system's weight is a reduction of cables and connectors: The MAIUS-A apparatus had many meters of separate copper wires for clock, data, address selection and trigger signals. The goal is to replace most of those cables by a single connection, which is easy to route through the setup.

    Especially in the \ac{beccal} project, which occupies three separately powered lockers on \ac{iss}, but also in lab-setups, the electrical ground potential of different electronic experiment parts may not be the same. Thus, a galvanic isolation in the communication system is useful to prevent ground level shifting. This can be achieved by using optical fibers instead of copper cables.

    For the \ac{bec} experiments a central trigger impulse is needed, which arrives at all endpoints in deterministic time. For repeatability those trigger signals require a jitter performance in the nanosecond scale and have to keep the same offsets even when power-cycling the experiment.

    All reconfigurable electronic parts of \ac{beccal} should be programmable while in orbit for flexibility and for the possibility of error mitigation. Therefore, the network must be capable of sending bitstreams or device images of multiple megabytes in size quickly, to reduce the experiment's downtime.

    Finally, the system is limited on using hardware which is low on power, thermal dissipation and fits the small budget. In this case only \ac{cots} and small form factor hardware may be used. As the endpoints will interface with highly specialized custom sensors and actuators at a high speed, \acp{fpga} are used, which will provide the performance needed for the communication stack while staying in the mentioned budgets.

    Possible candidates for the communication system are real-time Ethernet (e.g. Profinet or Ethernet POWERLINK), SpaceWire or SpaceFibre \cite{feld2004profinet,parkes2005spacewire,parkes2010spacefibre}. Profinet does not suffice the required timing constraints, does not prevent ground level shifts and a typical interface implementation is too large for smaller \ac{cots} \acp{fpga} \cite{ishak2013reducing,germanos2015synchronizing}.
    Another Ethernet-based candidate presented by Födisch et al. in 2016 has a resolution of \SI{8}{\nano\second} but again is too large and is not isolated \cite{fodisch2016synchronous}.
    While SpaceWire is also based on electrical connection, which does not fulfill the galvanic isolation requirement, its successor SpaceFibre seems promising. But as for real-time Ethernet, a typical interface implementation is also too large for smaller \acp{fpga}, especially when switches with multiple ports in the network are needed.
    Key points of those technologies are shown in \cref{tab:network-comparison} in \cref{sec:summary}.

    In \cref{sec:system-design} a proposed communication system which fulfills the requirements listed above will be described. The design choices to implement high synchronicity while maintaining a low complexity are explained in \cref{sec:system-synchronicity,sec:system-implementation-complexity}. Afterwards, the evaluation of the system's main goals is described and the results are discussed in \cref{sec:measure-eval}.

\section{System design} \label{sec:system-design}
    The proposed communication system has a tree topology and consists of a single host as the tree root, multiple switches for branching and endpoints as tree leaves. The communication between two hops of the network is realized via plastic optical fibers, one for each transmission direction: The fiber transmitting data and timing information from a master port to the slave port is called Tx fiber, the receiving one is called Rx fiber.
    The \emph{host} is the master of the whole network. Only master ports are implemented here, which supply a clock to the connected slave ports. There may be multiple hosts in a network for failover.
    \emph{Switches} have master and slave ports, which forward communication requests originating from a connected device. Multiple slave ports may be configured at a switch for redundancy.
    An \emph{endpoint} only has slave ports and is the termination point of communication in the network. Transported user data payloads will be processed there.

\begin{figure}[h!]
        \centering
        \scalebox{0.9}{
            \input{figs/network_topology}
        }
        \caption{Planned network topology for the BECCAL apparatus. Plastic optical fiber connections (in pairs of Tx and Rx fibers) run between the host, the switches and the endpoint cards. The network is controlled by a computer connected via Ethernet.}
        \label{fig:beccal-network-topology}
    \end{figure}
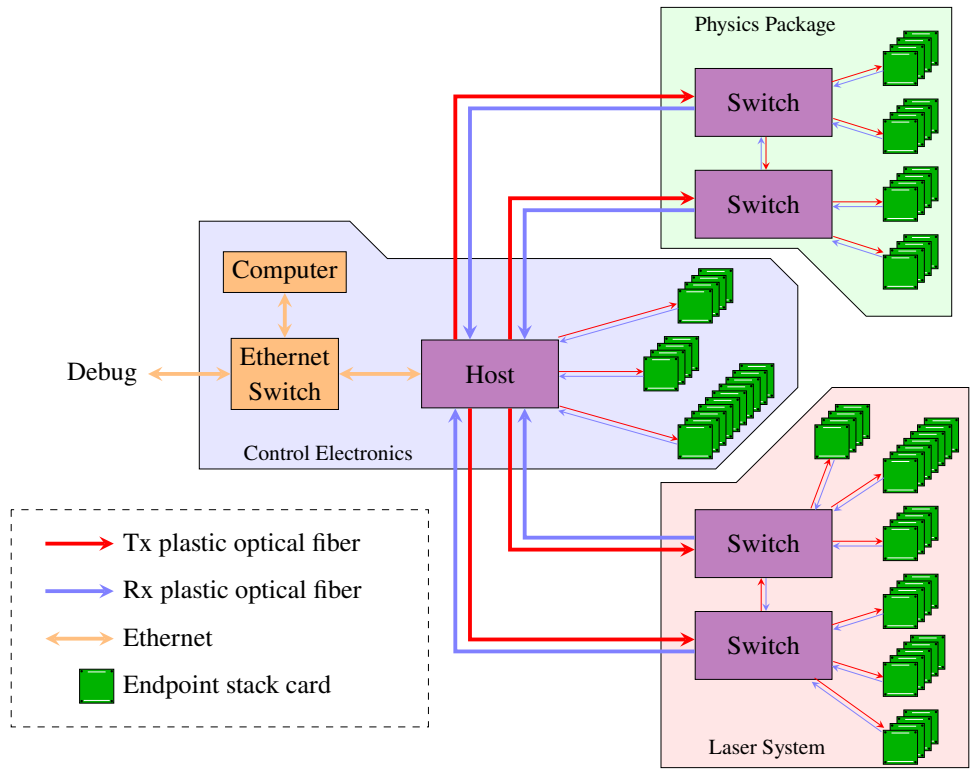

    \Cref{fig:beccal-network-topology} exemplarily shows the planned network for the \ac{beccal} project divided into the control electronics, the physics package and the laser system. The endpoints are electronic boards with a stackable PC/104-like connector. They drive sensors and actuators on boards stacked onto them by executing commands from the network, for example the laser phase-lock in the laser system or the electromagnets of magneto-optical traps in the physics system. The \ac{bec} experiments are controlled by a computer which is connected to the host by Ethernet.

    The use of a single plastic optical fiber pair instead of multiple wires reduces the weight and routing complexity and also provides galvanic isolation. In \cref{sec:system-synchronicity,sec:system-implementation-complexity} the design decisions are described to implement synchronization and maintaining a high data rate while using fewer resources.

    \subsection{Network synchronicity} \label{sec:system-synchronicity}
        One main goal of the network is to provide a method for precise deterministic global timing. The network should keep the same communication delays and behavior even if it is restarted. Therefore, deterministic synchronization of transmitted bits and symbols is needed.

\begin{figure}[h!]
            \centering
            \scalebox{0.95}{
                \input{figs/tx_fiber_example_setup}
            }
            \input{wave/tx_fiber_clock}
            \caption{Block and timing diagram of the Tx fiber bit level protocol for a transmission from a master to a slave. The slave's PLL locks onto the rising edges of the Tx fiber signal (first marked orange). The transmitted data bit sequence is 101, each actual data bit is marked by a circle and sampled on the falling edge of the recovered slave clock.}
            \label{fig:physical-layer-example}
        \end{figure}
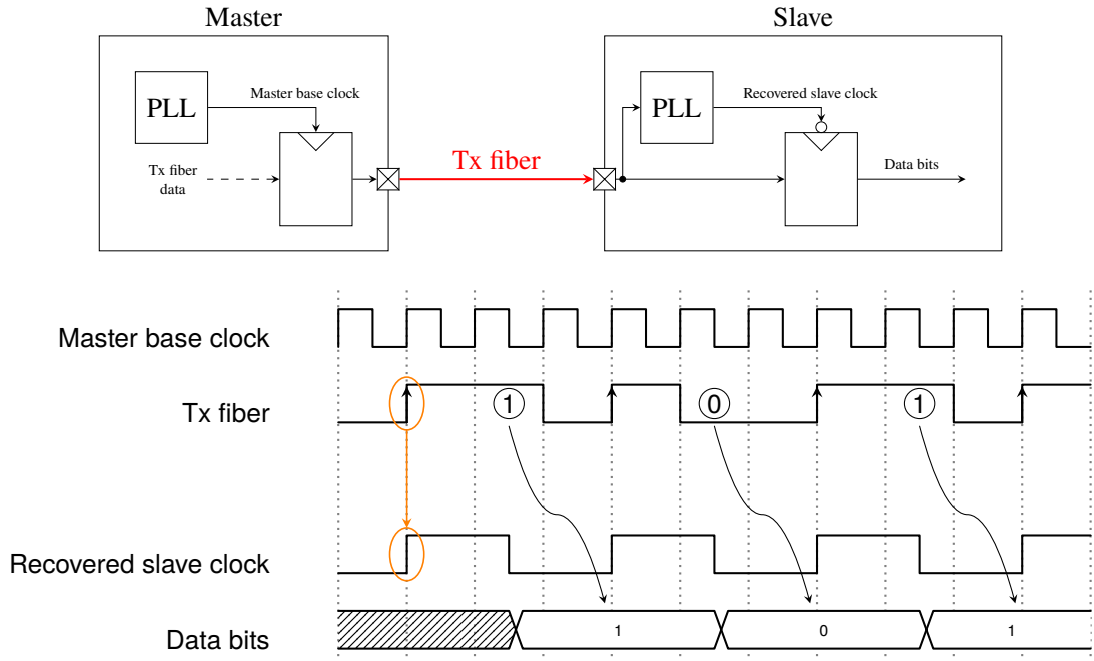

        Bit synchronization of the network is performed by using a protocol which encapsules data in a timing-recoverable signal for the Tx fiber. The used base clock is three times faster than the bit rate: Every data bit transmission takes three cycles, starting with a logical 1 (high) followed by the actual data bit and a logical 0 (low) for one cycle each. Thus, every third clock cycle a rising edge occurs, on which a \ac{pll} at a following slave port deterministically can lock onto. The slave can sample the data bits on the falling edges of that recovered clock. \Cref{fig:physical-layer-example} shows an example of the Tx fiber protocol.

        The Rx fiber protocol does not contain information for timing recovery. The data bits are transferred ``as-is'' on a base clock with a third of the frequency of the Tx fiber base clock. They can be decoded at the receiver by oversampling.

        Symbol synchronization is done by encoding every byte with the 8b/10b line code described by Widmer and Franaszek which adds a running disparity and multiple in-band signaling and synchronization words \cite{Widmer1983}. It redundantly encodes every data byte as 10\,bit symbols on the line. Some symbols not used for data bytes are used as control words. In the beginning and when a point-to-point connection does not transmit data, an idle symbol is transmitted. The special 8b/10b comma symbols' sequences of bits, K.28.1 and K.28.5, cannot occur as a part of another valid 8b/10b symbol stream (they are prefix-free) and thus can be used for synchronization by detecting the start of a valid symbol in the bitstream.

        Using the bit and symbol synchronization methods, a slave device can deterministically lock its clock and symbol processing onto a master device. At this point a trigger symbol can be inserted, which can be used for synchronous triggering at the endpoints: The trigger symbol is a special 8b/10b control symbol which can be sent at any time. When the host sends a trigger symbol, all switches will immediately forward it to their connected devices. As all point-to-point connections deterministically are bit and symbol synchronous with a constant phase shift and the trigger symbol is forwarded with a constant delay at switches, the latencies of a trigger symbol distributed from the host to all endpoints is constant, too. Those latencies can be measured beforehand and can be used to synchronize the trigger execution at endpoints by using high frequency counters.

    \subsection{Implementation complexity and transfer speeds} \label{sec:system-implementation-complexity}
\begin{figure}[h!]
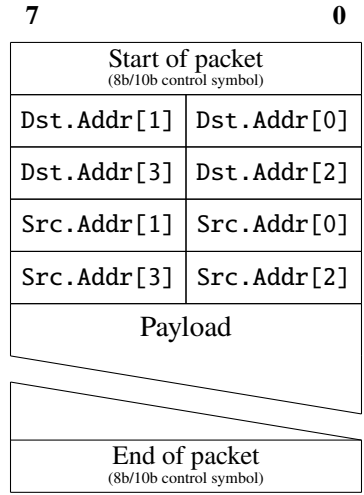

            \centering
            \begin{bytefield}[bitformatting={\small\bfseries}, bitwidth=1.5em, endianness=big]{8}
                \bitheader{7,0} \\
                \bitbox{8}{\small{Start of packet}\\\tiny{(8b/10b control symbol)}}\\
                \bitbox{4}{\small{\texttt{Dst.Addr[1]}}} 
                \bitbox{4}{\small{\texttt{Dst.Addr[0]}}}\\
                \bitbox{4}{\small{\texttt{Dst.Addr[3]}}} 
                \bitbox{4}{\small{\texttt{Dst.Addr[2]}}}\\
                \bitbox{4}{\small{\texttt{Src.Addr[1]}}} 
                \bitbox{4}{\small{\texttt{Src.Addr[0]}}}\\
                \bitbox{4}{\small{\texttt{Src.Addr[3]}}} 
                \bitbox{4}{\small{\texttt{Src.Addr[2]}}}\\
                \wordbox[lrt]{1}{Payload} \\
                \skippedwords \\
                \bitbox{8}{\small{End of packet}\\\tiny{(8b/10b control symbol)}}
            \end{bytefield}
            \caption{Transport layer data packet fields from bit 7 (MSB) down to bit 0. Start and end are indicated by 8b/10b control words apart the data stream.}
            \label{fig:data-packet}
        \end{figure}

        To provide high throughput and low resource usage in the implemented system, a simple data structure and addressing scheme should be defined which does not need complex routing structures or address tables.
        The user data is embedded in packets which contain the destination address by which they are routed to their destination. A packet begins with the transmission of a \ac{sop} 8b/10b control symbol, followed by a packet header which contains fields for the destination and source address. Afterwards the payload is transferred and the packet finished with sending the 8b/10b control symbol \ac{eop}. By design the packets are not limited in size. \Cref{fig:data-packet} shows the fields of a data packet.

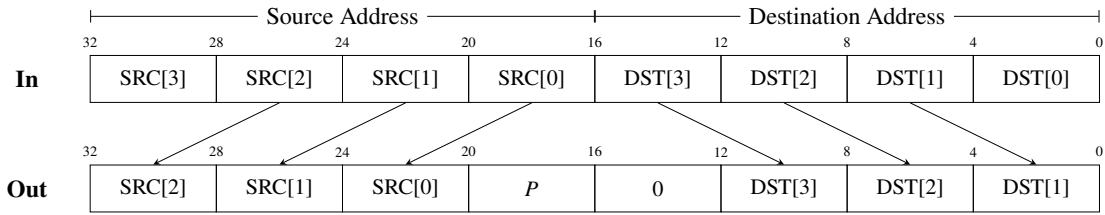
\begin{figure}[ht]
          \centering
          \scalebox{0.83}{
            \input{figs/address_shift_example}
          }
          \caption{Address part modification at an intermediate device. The packet is coming in on port $P$. The destination addresses are shifted to the right by one entry, the incoming port number is added to the source address on the right.}
          \label{fig:address_shift_example}
        \end{figure}

        The destination address nibble (4\,bit) fields are simply the port numbers at the host or switches to which the packet should be forwarded. Every switch will forward the packet to the port in \texttt{Dst.Addr[0]} with the destination fields shifted to the right by one nibble, removing the used part and adding four zero bits at \texttt{Dst.Addr[3]}. When a packet arrives with a 0 for \texttt{Dst.Addr[0]} its destination has been reached and the device can process the contents. Every device in the network of up to 4 hops and with up to 14 slave devices (one address is the master port, and address 0 is the local device) can be addressed that way.
        Simultaneously the source address fields at every hop are filled with the port where the packet arrived to provide a reply address for the packet. The shifting is shown in \cref{fig:address_shift_example}.

        With this addressing scheme any intermediate device only needs a buffer of 2 Bytes for the destination address fields to initially determine the target port and shift the address. After adding the source port to the shifted source address all other data just can be forwarded. No address tables are needed, which reduces the need of complex logic constructs. Only the host needs to know the network's topology for sending packets as the reply address of a packet is generated on the fly when traversing the network. Because the packets do not need to be buffered at every hop and can be passed through after 2 Bytes, low delays and high data rates are possible.

        For flow control two different 8b/10b idle symbols can be sent: One to signal the availability of buffer capacity (ready to receive) and another to signal that the buffer is running full (not ready to receive). Those idle symbols propagate back through the network to momentary stop transmission of new data. 
        A hysteresis on the buffer levels can be used to block or unblock the connection. The buffer should only be used to temporarily store data while the network is throttled down automatically by flow control and usually data should be forwarded immediately. Therefore, only several bytes need to be added as buffers at intermediate devices.

\section{Measurement and evaluation} \label{sec:measure-eval}
\begin{figure}[h!]
        \centering
        \begin{tikzpicture}[font=\footnotesize]
            \begin{pgfonlayer}{background}
            \node[anchor=south west,inner sep=0] (image) at (0,0)
            {\includegraphics[width=.8\textwidth]{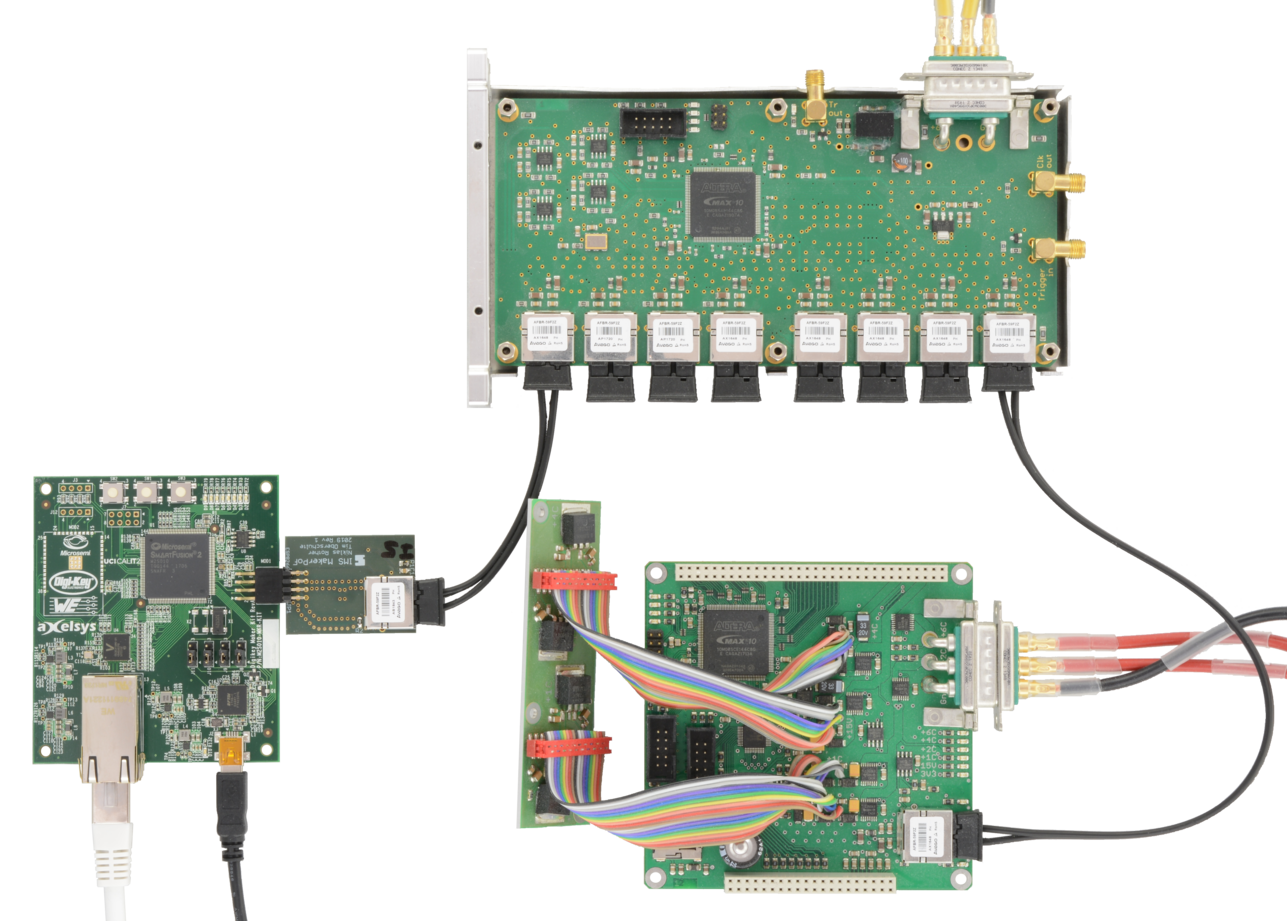}};
            \end{pgfonlayer}
            \begin{scope}[x={(image.south east)},y={(image.north west)}]
                \node[shape=circle,draw,inner sep=2pt] at (0.1228,1-0.47) {A};
                \node[shape=circle,draw,inner sep=2pt] at (0.5820,1-0.0565) {B};
                \node[shape=circle,draw,inner sep=2pt] at (0.6325,1-0.5698) {C};
            \end{scope}
        \end{tikzpicture}
        
        \caption{Photograph of the setup with a host \Circled{A}, switch \Circled{B} and an endpoint \Circled{C}. The host is connected with the control computer via Ethernet and powered via USB. Its fiber master port is connected to the upstream (slave) port of the switch. The switch is powered via the cable on its top right and connects over the rightmost fiber master port to the endpoint.}
        \label{fig:picture-setup}
    \end{figure}
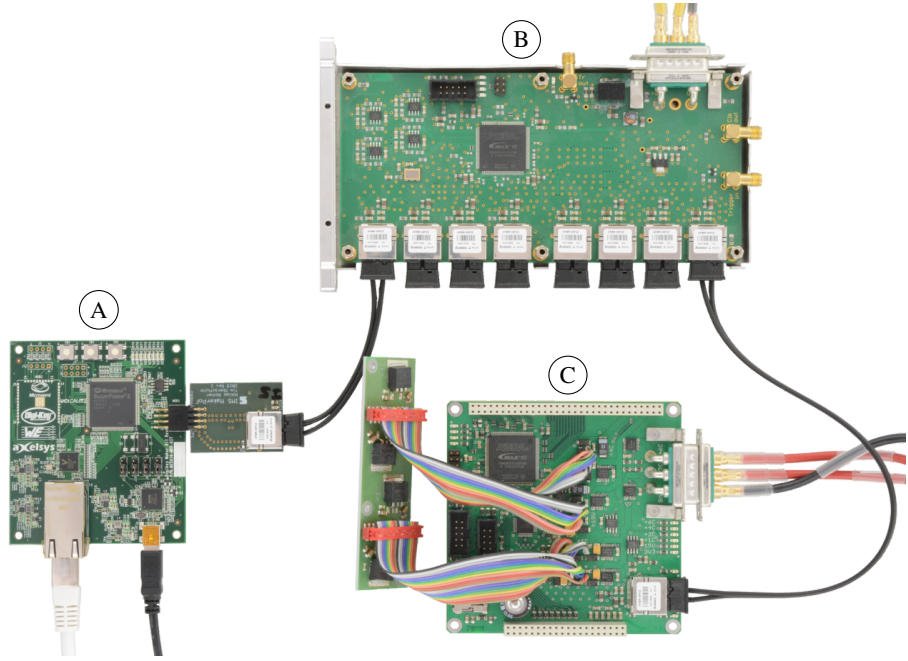

    The system described in \cref{sec:system-design} was implemented using Intel MAX10 and Microchip SmartFusion2 \acp{fpga}.
    The circuit boards for a host device, a switch and an endpoint were designed and the \ac{fpga} firmware for each was implemented.
    The hardware is shown in \cref{fig:picture-setup}.
    
    The host device uses a Microchip SmartFusion2 \ac{soc} and functions as a bridge from Ethernet into the fiber network. Its ARM Cortex-M3 hard processing system is used for Ethernet and TCP operations and packet processing. The \ac{fpga} fabric is used for the fiber protocol's transport layer encoding and hardware interface access. It generates the main clock signal for the network and can send global triggers. It is based on an evaluation board for prototyping purposes.

    The switch is implemented on an Intel MAX10 \ac{fpga} and is configured to have one upstream (slave) port, where it recovers its clock from, and seven downstream (master) ports where other switches or endpoints can be connected. The internal configuration of the switch can be accessed by a virtual interface at port 0 (the local device address) which makes the switch usable like an endpoint. It also contains an SMA connector trigger output for debugging.

    The endpoint implementation is also implemented on an Intel MAX10 \ac{fpga}. This endpoint has a configurable trigger unit for precisely triggering cards stacked onto it via the two PC/104-like connectors which contain 10 separate trigger lines.

    \subsection{Weight and complexity reduction}
        The former interconnect used in the MAIUS-A mission contained one Ethernet cable for data, 10 copper cables for trigger lanes, and one coaxial cable for the clock signal per endpoint. All of those will be replaced by the presented network system with only one pair of plastic fibers for each hop. In addition to the reduced routing complexity inside the experiment it also has an impact onto the weight: The 10 formerly used No. 28 AWG copper wires in a ribbon cable with a cross-sectional area of \SI{0.0804}{\milli\meter\squared} had a weight of \SI{3}{\gram\per\meter}, the RG-174 coaxial cable a weight of \SI{12}{\gram\per\meter} and the CAT6 Ethernet cable added \SI{46}{\gram\per\meter}. The total interconnection mass with copper wires calculates as
        \[ 10 \times \SI{3}{\gram\per\meter} + \SI{12}{\gram\per\meter} + \SI{46}{\gram\per\meter} = \SI{88}{\gram\per\meter}. \]
        Using two plastic optical fibers, with a mass of \SI{3.8}{\gram\per\meter} each, the total interconnection mass will be \SI{7.6}{\gram\per\meter}, a reduction by more than a factor of 10.

        For the BECCAL mission with a total interconnect length of approximately \SI{23}{\meter} this means a reduction by
        \[ \SI{23}{\meter} \times \SI{88}{\gram\per\meter} - \SI{23}{\meter} \times \SI{7.6}{\gram\per\meter} = \SI{2.024}{\kilo\gram} - \SI{0.1748}{\kilo\gram} \approx \SI{1.85}{\kilo\gram} \]
        when using the proposed fiber network instead of the copper wires.

    \subsection{Trigger delay variance}
        The deterministic synchronization of the system, especially of the trigger signal, is of capital importance. Therefore, the jitters of the clock and the trigger delays were measured.

        First, the constancy of the locking onto the clock signal by a slave with the same delay was evaluated. In case of the host and the switch, the rising edges of the host's oscillator and the slave's recovered clock from the Tx line are \SI{7.58}{\nano\second} apart with a standard deviation of \SI{46}{\pico\second}. This also holds when power-cycling any of the devices, i.e. resynchronizing them. As the host's oscillator already jitters with a standard deviation of \SI{36}{\pico\second}, this variance could be reduced drastically by using a more stable oscillator at the host.

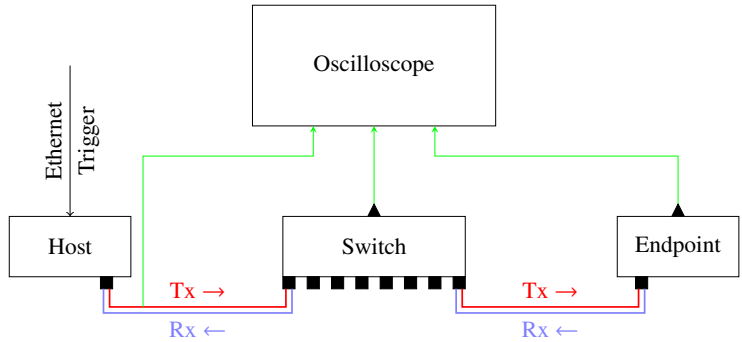
\begin{figure}[h!]
            \centering
            \scalebox{0.8}{
                \input{figs/measurement_setup}
            }
            \caption{Setup of the trigger delay measurement. The fiber ports are depicted as squares, the trigger outputs as triangles. The devices are connected via the Tx and Rx plastic fiber pairs. On the oscilloscope the triggers of switch and endpoint are connected directly whilst the host's Tx fiber signal is captured to detect the trigger symbol. The trigger is started by a special Ethernet frame.}
            \label{fig:trigger-delay-setup}
        \end{figure}

\begin{figure}[h!]
            \centering
            \includegraphics[width=\textwidth]{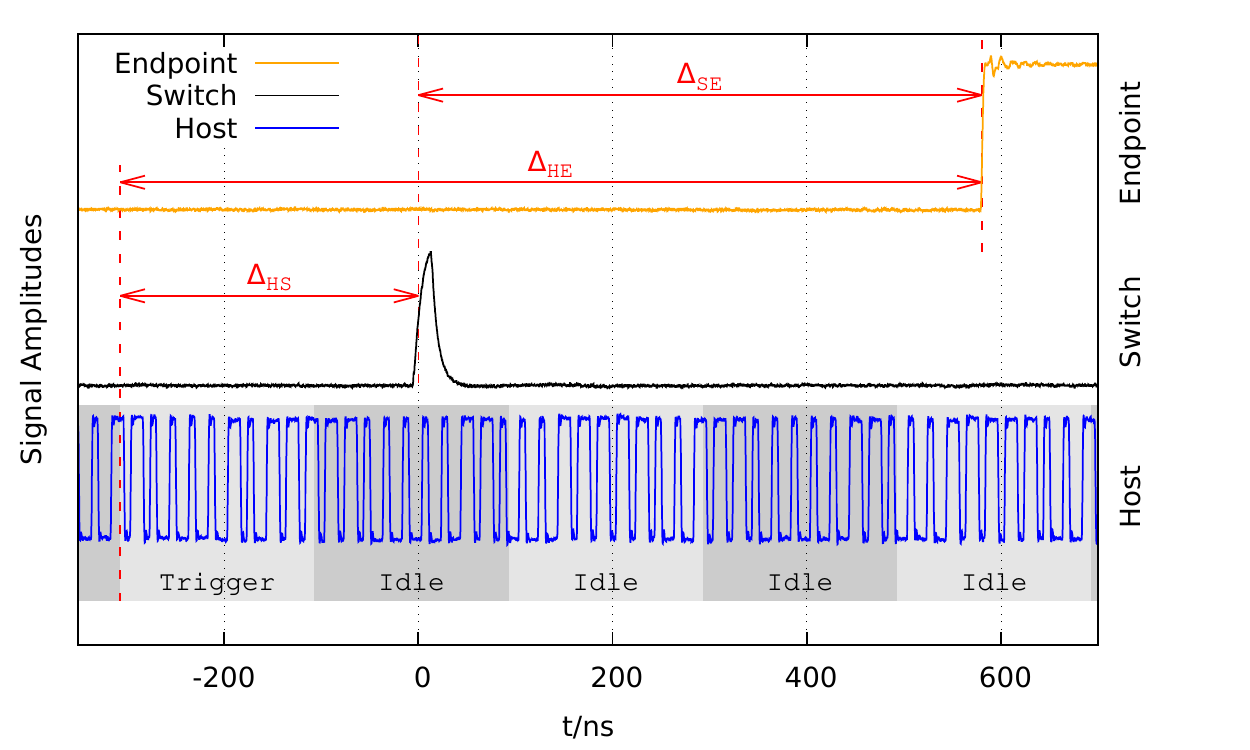}
            \caption{Exemplary delay measurement of one trigger distribution. The bottom graph shows the 8b/10b Tx signal of the host encoded as explained in \cref{fig:physical-layer-example}, including the 8b/10b trigger signal and multiple idle signals. The middle graph displays the measured trigger output of the switch, the top one shows the trigger outputted by the endpoint.}
            \label{fig:trigger-delay-measurement}
        \end{figure}

        To measure the board-to-board trigger jitter, the global trigger signal sent by the host is outputted on an SMA connector at the switch and at the endpoints. Thus, the variances of the delays of the trigger signals between the cards were measured and analyzed as depicted in \cref{fig:trigger-delay-setup}. The trigger has been sent from the host into the network \num{5000} times and each one has been recorded by an oscilloscope at a sample rate of \SI{5}{\giga\hertz}. As reference points for the measurement the beginning of the 8b/10b trigger symbol and rising edges (\SI{50}{\percent} low to high) of the trigger outputs at the switch and endpoint were used. An exemplary measurement is shown in \cref{fig:trigger-delay-measurement}. $\Delta_\text{HS}$, $\Delta_\text{HE}$, and $\Delta_\text{SE}$ denote the evaluated delays from host to switch, from host to endpoint, and from switch to endpoint, respectively.

        \begin{table}[h!]
            \centering
            \caption{Measured variances of the trigger occurrences as shown in \cref{fig:trigger-delay-measurement}. SD is the standard deviation.}
            \begin{tabular}{ll|r|r|r|r}
                & Delay               & Mean (\si{\nano\second})  & Min (\si{\nano\second})   & Max (\si{\nano\second})   & SD (\emph{\si[detect-weight=true]{\pico\second}}) \\ \hline
                Host to Switch & $\Delta_\text{HS}$  & 319.46                    & 318.72                    & 320.32                    & 248.2  \\
                Switch to Endpoint & $\Delta_\text{SE}$  & 581.26                    & 580.45                    & 582.08                    & 239.2  \\
                Host to Endpoint & $\Delta_\text{HE}$  & 900.72                    & 900.30                    & 901.03                    & 99.6   \\
            \end{tabular}
            \label{tab:trigger-delay-results-single}
        \end{table}

        \Cref{tab:trigger-delay-results-single} shows that the trigger arrival times in the test network are distributed with a standard deviation of maximum \SI{248.2}{\pico\second} and in a range of \SI{1.63}{\nano\second}. Despite those high variances of the two segments $\Delta_\text{HS}$ and $\Delta_\text{SE}$, when looking at the measures for the chained Host to Endpoint delay $\Delta_\text{HE}$, its variance counterintuitively is smaller with a standard deviation of \SI{99.6}{\pico\second} in a range of \SI{0.73}{\nano\second}. That is most likely due to the switch trigger output having a higher jitter because the cable is longer, affecting the measures from and to the switch. However, this output discrepancy does not affect the full chain variance. The same results were achieved when power cycling the network between the trigger transmissions.

    \subsection{Network data rates}
        The theoretical data rate of the channel is limited by the clock frequency and the number of data bits transferred at a time. The used Tx base clock is \SI{150}{\mega\hertz}, but only every third bit is a data bit, resulting in one bit of the 8b/10b code with a rate of \SI{50}{\mega\hertz}. In this code \SI{8}{\bit} of data are encoded as \SI{10}{\bit} on the channel, limiting the data rate to \SI[per-mode=repeated-symbol]{40}{\mega\bit\per\second}.

        To measure the achievable data rate, packets of multiple sizes were transferred over the channel. At the switch and the endpoint parts or multiples of an existing 256 Byte firmware ROM were read via a custom application layer protocol for the data transfers. All data in the payload of the transport layer packets as shown in \cref{fig:data-packet} is counted as user data for the measurements, i.e. the address and start/end symbols are excluded.

        \begin{figure}[h!]
            \centering
            \includegraphics[width=.85\textwidth]{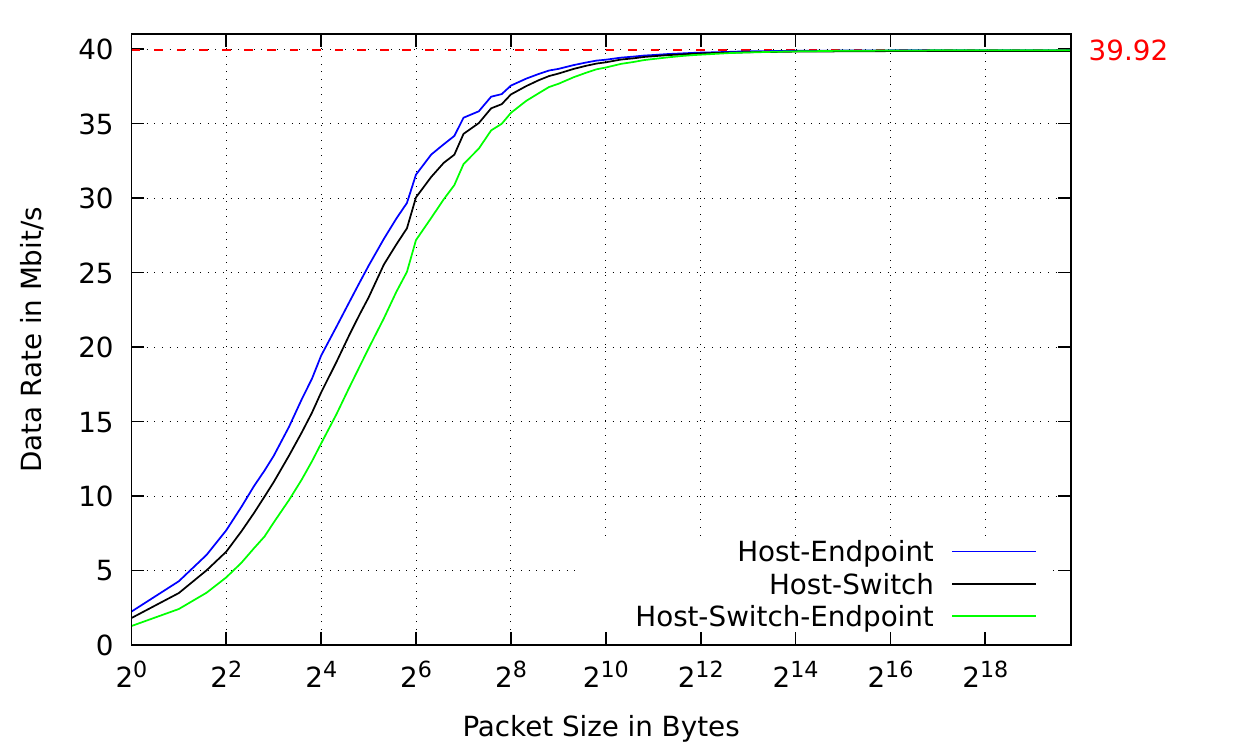}
            \caption{Median data rate measured when reading data directly from an endpoint, a switch and an endpoint via a switch over the proposed network for multiple packet user data sizes (displayed on a logarithmic scale). The dashed line marks the maximum rate of \SI{39.92}{\mega\bit\per\second} reached with the biggest packets.}
            \label{fig:data-rate-results}
        \end{figure}

        \Cref{fig:data-rate-results} shows the results of the measurements for the data rates from the host directly to the endpoint, to the switch and to the endpoint routed through the switch. For smaller packets the data rate is lower because the overhead of control data sent with every packet is bigger than the user data itself. When increasing the packet sizes, the data rate increases up to a maximum of \SI{39.92}{\mega\bit\per\second} at a packet size of about \SI{918}{\kilo\byte}. The small discontinuities in the mid-range of the diagram are caused by reaching the maximum buffer sizes in the host and resulting waiting cycles.

        The switches data rate is a bit lower than the one of the endpoint, because the implementations slightly differ. The switch has two more queues in the switching fabric which delay a packet by multiple hundred nanoseconds. Furthermore the chain of switch and endpoint has the highest delivery time and therefore the rise in data rate at higher packet sizes compared to the other two.

        A higher data rate may be reached when increasing the clock frequencies. This is limited by the capabilities of the ports of the \acp{fpga} and the fiber port switching capabilities.

    \subsection{Resource usage}
        All used \acp{fpga} have a low power footprint with limited fabric capabilities of only \SI{8}{\kilo\logicelements} (\aclp{le}) to \SI{10}{\kilo\logicelements}. The \ac{fpga} firmware was implemented in VHDL without using vendor specific components (with exception of the PLLs) and is portable to \acp{fpga} of other manufacturers.
        Each fiber interface, including the bit-level and 8b/10b-level encoding and decoding units, uses about \SI{590}{\logicelements} (LUT-4) and 2 memory blocks. This is less then one-tenth of the size of the SpaceFibre interface implementation \cite{7500644}. The 8 port switch additionally uses units for packet arbitration, identification and monitoring, resulting in \SI{6438}{\logicelements} for the whole \ac{fpga} design and fits the Intel MAX10 10M08 \ac{fpga}.

\section{Summary} \label{sec:summary}
    It is shown that it is possible to build a low-cost, deterministic low-jitter isochronous and galvanically isolated network with a simple protocol and a data rate of almost \SI{40}{\mega\bit\per\second} for experimental setups and realize a proof-of-concept implementation. Deterministic global network synchronicity precision of under \SI{1}{\nano\second} with a standard deviation of less than \SI{100}{\pico\second} is possible when only using commercial parts. \Cref{tab:network-comparison} shows the proposed network in comparison to some existing technologies.

    \begin{table}[h!]
        \caption{Comparison of existing network technology in terms of connection type, timing resolution, typical interface implementation size on \acp{fpga} in number of \acp{le} and data rate.}
        \centering
        \begin{tabular}{l|r|r|r|r}
            Name                            & Connection    & Resolution            & Size                              & Data rate\\ \hline
            Profinet \hfill \cite{ishak2013reducing,germanos2015synchronizing}
                & Electrical    & \SI{1}{\milli\second} & \SI{7}{\kilo\logicelements}       & \SI{100}{\mega\bit\per\second}\\
            SpaceWire \hfill \cite{7500644,van2011spacewire,mcclements2003spacewire}
                & Electrical    & \SI{7}{\micro\second} & \SI{0.52}{\kilo\logicelements}    & \SI{200}{\mega\bit\per\second}\\
            SpaceFibre \hfill \cite{7500644,10.1007/978-981-15-4163-6_36}
                & Optical       & \SI{58}{\nano\second} & \SI{7.2}{\kilo\logicelements}     & \SI{2000}{\mega\bit\per\second}\\
            Ethernet \emph{Födsch et al.} \hfill \cite{fodisch2016synchronous}
                & Electrical    & \SI{8}{\nano\second}  & \SI{3.4}{\kilo\logicelements}     & \SI{1000}{\mega\bit\per\second}\\
            \emph{proposed network}      
                & Optical       & $<\SI{1}{\nano\second}$& \SI{0.59}{\kilo\logicelements}   & \SI{40}{\mega\bit\per\second}\\
        \end{tabular}
        \label{tab:network-comparison}
    \end{table}

    Furthermore, the goal of reducing the weight and complexity of the network used in the MAIUS-A and BECCAL missions is reached by sending clock, trigger and data signals over a single plastic fiber. This reduces the weight by more than a factor of 10 and cuts down the number of cables from 3 to 1 for every connection.

    Aside from the space missions this network can also be used for custom laboratory setups where high synchronicity at a low cost is needed. Its galvanic isolation and fiber lengths of up to \SI{50}{\meter} also makes it usable when spanning multiple rooms with different electrical grids.

    \bibliography{bibliography}

\section*{Acronyms}
    \begin{acronym}
        \acro{bec}[BEC]{Bose-Einstein condensate}
        \acro{beccal}[BECCAL]{Bose-Einstein Condensate Cold Atom Laboratory}
        \acro{cots}[COTS]{commercial off-the-shelf}
        \acro{eop}[EOP]{End of Packet}
        \acro{fpga}[FPGA]{Field Programmable Gate Array}
        \acro{iss}[ISS]{International Space Station}
        \acro{le}[LE]{Logic Element}
        \acro{pll}[PLL]{phase-locked loop}
        \acro{soc}[SoC]{System on a Chip}
        \acro{sop}[SOP]{Start of Packet}
    \end{acronym}

\end{document}

%% file: figs/network_topology.tex
\newcommand\DoubleLine[5][3pt]{%
  \path(#2)--(#3)coordinate[at start](h1)coordinate[at end](h2);
  \draw[#4]($(h1)!#1!90:(h2)$)--($(h2)!#1!-90:(h1)$);
  \draw[#5]($(h1)!#1!-90:(h2)$)--($(h2)!#1!90:(h1)$);
}

\begin{tikzpicture}[
    tbusboard/.pic={
    	\node (-inner) at (0.0,0.0) [draw,fill=black!30!green,minimum width=0.5cm,minimum height=0.5cm] {};%
			\draw [white] 	(-0.13,0.21) -- (0.13,0.21);
			\draw [white] 	(-0.19,-0.21) -- (0.19,-0.21);
			\draw 			(-0.22,-0.22) circle (0.01);
			\draw 			(0.22,-0.22) circle (0.01);
			\draw 			(-0.22,0.22) circle (0.01);
			\draw 			(0.22,0.22) circle (0.01);
		},
		/tikz/tbusstack/board count/.initial={2},
		tbusstack/.pic={
			\pgfkeysgetvalue{/tikz/tbusstack/board count}{\boardcnt}
    	\foreach \x in {{\boardcnt},...,0} {\path (0.\x,0.\x) pic (-board-\x) {tbusboard};};
		},
	]

	%% Regions
	\draw [fill=blue!10] (-1.5,1.7) -- (4.1,1.7) -- (4.5,1.3) -- (4.5,-0.2) -- (3.3,-1.4) -- node[align=center, above, pos=0.75, font=\footnotesize] {Control Electronics} (-4.25,-1.4) -- (-4.25,2.25) -- (-2.05,2.25) -- cycle;

	\draw [fill=red!10] (2.5,-1.6) -- (3.6,-1.6) -- (5,-0.2) -- (7,-0.2) -- (7,-5.8) -- node[align=center, above, xshift=-0.7cm, font=\footnotesize] {Laser System} (2.5,-5.8) -- cycle;

	\draw [fill=green!10] (2.5,1.85) -- (4.5,1.85) -- (5.5,0.85) -- (6.75,0.85) -- (6.75,5.4) -- node[align=center, below, xshift=-0.6cm, font=\footnotesize] {Physics Package} (2.5,5.4) -- cycle;

	%% Nodes
	\node (eth-switch) at (-3,0) [draw,align=center,fill=orange!50] {Ethernet\\Switch};
	\node (computer) at (-3,1.5) [draw,align=center,fill=orange!50] {Computer};

	\node (main-switch) at (0,0) [draw,align=center,fill=violet!50,minimum height=1cm, minimum width=2cm] {Host};
	\node (pp-switch)   at (4,2.5) [draw,align=center,fill=violet!50,minimum height=1cm, minimum width=2cm] {Switch};
	\node (pp-switch2)  at (4,4) [draw,align=center,fill=violet!50,minimum height=1cm, minimum width=2cm] {Switch};
	\node (ls-switch1)  at (4,-2.5) [draw,align=center,fill=violet!50,minimum height=1cm, minimum width=2cm] {Switch};
	\node (ls-switch2)  at (4,-4) [draw,align=center,fill=violet!50,minimum height=1cm, minimum width=2cm] {Switch};

	\draw pic[tbusstack/board count=3] (pp-tbus4) at (6,4.5) {tbusstack};
	\draw pic[tbusstack/board count=3] (pp-tbus1) at (6,3.5) {tbusstack};
	\draw pic[tbusstack/board count=3] (pp-tbus2) at (6,2.5) {tbusstack};
	\draw pic[tbusstack/board count=3] (pp-tbus3) at (6,1.5) {tbusstack};

	\draw pic[tbusstack/board count=3] (ce-tbus1) at (3,1) {tbusstack};
	\draw pic[tbusstack/board count=3] (ce-tbus2) at (2.5,0) {tbusstack};
	\draw pic[tbusstack/board count=9] (ce-tbus3) at (3,-1) {tbusstack};

	\draw pic[tbusstack/board count=3] (ls1-tbus1) at (5,-1) {tbusstack};
	\draw pic[tbusstack/board count=6] (ls1-tbus2) at (6,-1.5) {tbusstack};
	\draw pic[tbusstack/board count=3] (ls1-tbus3) at (6,-2.5) {tbusstack};

	\draw pic[tbusstack/board count=3] (ls2-tbus1) at (6,-3.5) {tbusstack};
	\draw pic[tbusstack/board count=4] (ls2-tbus2) at (6,-4.5) {tbusstack};
	\draw pic[tbusstack/board count=3] (ls2-tbus3) at (6,-5.5) {tbusstack};

	%% Interconnect
	\draw [ultra thick,red, -stealth] (main-switch.135) |- (pp-switch2.175);
	\draw [ultra thick,blue!50,   stealth-] (main-switch.120) |- (pp-switch2.185);

	\draw [ultra thick,red, -stealth] (main-switch.60)  |- (pp-switch.175);
	\draw [ultra thick,blue!50,   stealth-] (main-switch.45)  |- (pp-switch.185);

	\draw [ultra thick,red, -stealth] (main-switch.240) |- (ls-switch2.175);
	\draw [ultra thick,blue!50,   stealth-] (main-switch.225) |- (ls-switch2.185);

	\draw [ultra thick,red, -stealth] (main-switch.300) |- (ls-switch1.185);
	\draw [ultra thick,blue!50,   stealth-] (main-switch.315) |- (ls-switch1.175);

	\DoubleLine[1pt]{pp-switch2.-15}{pp-tbus1-board-0-inner.west}{-stealth,red}{stealth-,blue!50}
	\DoubleLine[1pt]{pp-switch2.15}{pp-tbus4-board-0-inner.west}{-stealth,red}{stealth-,blue!50}
	\DoubleLine[1pt]{pp-switch.east}{pp-tbus2-board-0-inner.west}{-stealth,red}{stealth-,blue!50}
	\DoubleLine[1pt]{pp-switch.south east}{pp-tbus3-board-0-inner.north west}{-stealth,red}{stealth-,blue!50}

	\DoubleLine[1pt]{main-switch.north east}{ce-tbus1-board-0-inner.west}{-stealth,red}{stealth-,blue!50}
	\DoubleLine[1pt]{main-switch.east}{ce-tbus2-board-0-inner.west}{-stealth,red}{stealth-,blue!50}
	\DoubleLine[1pt]{main-switch.south east}{ce-tbus3-board-0-inner.west}{-stealth,red}{stealth-,blue!50}

	\DoubleLine[1pt]{ls-switch1.35}{ls1-tbus1-board-0-inner.south}{-stealth,red}{stealth-,blue!50}
	\DoubleLine[1pt]{ls-switch1.north east}{ls1-tbus2-board-0-inner.west}{-stealth,red}{stealth-,blue!50}
	\DoubleLine[1pt]{ls-switch1.east}{ls1-tbus3-board-0-inner.west}{-stealth,red}{stealth-,blue!50}

	\DoubleLine[1pt]{ls-switch2.15}{ls2-tbus1-board-0-inner.west}{-stealth,red}{stealth-,blue!50}
	\DoubleLine[1pt]{ls-switch2.-15}{ls2-tbus2-board-0-inner.west}{-stealth,red}{stealth-,blue!50}
	\DoubleLine[1pt]{ls-switch2.-35}{ls2-tbus3-board-0-inner.north west}{-stealth,red}{stealth-,blue!50}

	% Redundancy
	\DoubleLine[1pt]{ls-switch2.north}{ls-switch1.south}{-stealth,red}{stealth-,blue!50}
	\DoubleLine[1pt]{pp-switch2.south}{pp-switch.north}{-stealth,red}{stealth-,blue!50}

	\draw [ultra thick,orange!50, stealth-stealth] (main-switch) -- (eth-switch);
	\draw [ultra thick,orange!50, stealth-stealth] (eth-switch) -- (computer);
	\draw [ultra thick,orange!50, stealth-stealth] (eth-switch) -- ++(-2,0) node[left,black] {Debug};

	% %% Legend
	\draw [dashed] (-7,-2) rectangle (-0.9,-5.2);
	\draw [ultra thick,red,-stealth] (-6.5,-2.5) -- ++(1,0);
	\node at (-5.5,-2.5) [right,black] {{\strut}Tx plastic optical fiber};
	\draw [ultra thick,blue!50,-stealth] (-6.5,-3.2) -- ++(1,0);
	\node at (-5.5,-3.2) [right,black] {{\strut}Rx plastic optical fiber};
	\draw [ultra thick,orange!50,stealth-stealth] (-6.5,-3.9) -- ++(1,0);
	\node at (-5.5,-3.9) [right,black] {{\strut}Ethernet};
	\draw pic (legend) at (-5.75,-4.6) {tbusboard};
	\node at (-5.5,-4.6) [right,black] {{\strut}Endpoint stack card};

\end{tikzpicture}

%% file: figs/tx_fiber_example_setup.tex
\begin{tikzpicture}

    %Master
    \draw (0,1) |- node[pos=0.75,above]{Master} (4,4) |- cycle;
    \node[draw,minimum height=1cm, minimum width=1cm] (M_PLL) at (1,3) {PLL};
    \node[draw,minimum height=1.3cm, minimum width=1cm] (M_DAT) at (3,2) {};
    \draw (2.75,2.65) -- (3,2.4) -- (3.25,2.65);

    \draw[-stealth] (M_PLL) -| node[pos=0.45,above,font=\tiny] {Master base clock} (M_DAT);
    \draw[dashed, stealth-] (M_DAT.west) -- ++(-1.0,0) node[left,font=\tiny,align=center] {Tx fiber\\data};

    %Slave
    \draw (7,1) |- node[pos=0.75,above]{Slave} (12.5,4) |- cycle;
    \node[draw,minimum height=1cm, minimum width=1cm] (S_PLL) at (8,3) {PLL};
    \node[draw,minimum height=1.3cm, minimum width=1cm] (S_DAT) at (10,2) {};
    \draw (9.75,2.65) -- (10,2.4) -- (10.25,2.65);
    \draw (10,2.725) circle (0.075);

    \draw[-stealth] (S_PLL) -| node[pos=0.45,above,font=\tiny] {Recovered slave clock} (10,2.8);

    \draw[-stealth] (S_DAT) -- node[above,font=\tiny]{Data bits} ++(2,0);

    % Connection by fiber
    \node[fill=white,minimum height=0.3cm, minimum width=0.3cm,draw] (M_OUT) at (4,2) {};
    \draw (M_OUT.north west) -- (M_OUT.south east) (M_OUT.north east) -- (M_OUT.south west);
    \node[fill=white,minimum height=0.3cm, minimum width=0.3cm,draw] (S_IN)  at (7,2) {};
    \draw (S_IN.north west) -- (S_IN.south east) (S_IN.north east) -- (S_IN.south west);

    \draw[-stealth] (M_DAT) -- (M_OUT);
    \draw[thick,red,-stealth] (M_OUT) -- node[above]{Tx fiber} (S_IN);
    \draw[-stealth] (S_IN) -- (S_DAT);

    \draw[fill] (7.25,2) circle (0.04);
    \draw[-stealth] (7.25,2) |- (S_PLL);

    %%placeholder path
    \path (0,0.5) -- (1,0.5);

\end{tikzpicture}

%% file: wave/tx_fiber_clock.tex
\begin{tikztimingtable}[
    timing/yunit=0.5cm,
    timing/xunit=0.9cm,
    timing/c/arrow tip=stealth,
    timing/d/background/.style={fill=white},
    timing/u/background/.style={pattern=north east lines,pattern color=black}
    ]
    {Master base clock}         & [C] 22{0.5C}  \\
    {Tx fiber}                  & [timing/c/rising arrows] L N(mclk1) 1c 1.5H N(dat1) L N(mclk2) 1c 0.5H L N(dat2) L N(mclk3) 1c 1.5H N(dat3) L N(mclk4) 1c 0.5H  \\
    {}                          &   \\
    {Recovered slave clock}     & L N(sclk1) 6{1.5C} C \\
    {Data bits}                 & U U 0.5U N(result1) 3D{1} N(result2) 3D{0} N(result3) 2.5D{1} \\
    \extracode
    \begin{background}[help lines]
        \vertlines[dotted,thick]{}
    \end{background}

    \draw ($(dat1.MID) + (-0.5,0)$) circle (0.2cm) node{1};
    \draw[-stealth] ($(dat1.MID) + (-0.5,0)$) ++(0,-0.6) parabola [bend pos=0.5] ($ (result1.MID) + (1.4,0.7) $);

    \draw ($(dat2.MID) + (-0.5,0)$) circle (0.2cm) node{0};
    \draw[-stealth] ($(dat2.MID) + (-0.5,0)$) ++(0,-0.6) parabola [bend pos=0.5] ($ (result2.MID) + (1.4,0.7) $);

    \draw ($(dat3.MID) + (-0.5,0)$) circle (0.2cm) node{1};
    \draw[-stealth] ($(dat3.MID) + (-0.5,0)$) ++(0,-0.6) parabola [bend pos=0.5] ($ (result3.MID) + (1.4,0.7) $);

    \draw [orange,semithick]
         (mclk1.mid) ellipse (.25 and .7)
         (sclk1.mid) ellipse (.25 and .7);

    \draw[orange,semithick,-stealth]($ (mclk1.mid) - (0,.7) $) -- ($ (sclk1.mid) + (0,.7) $);

\end{tikztimingtable}

%% file: figs/address_shift_example.tex
\begin{tikzpicture}[scale=0.5]
  \node at (8,2.5) (srcText) {Source Address};
  \node at (16+8,2.5) (dstText) {Destination Address};

  \draw[|-] (0,2.5) -- (srcText);
  \draw[-|] (srcText) -- (16,2.5);

  \draw[|-] (16,2.5) -- (dstText);
  \draw[-|] (dstText) -- (32,2.5);

  \draw (0, 1.25) node[above] {\tiny $32$};
  \draw (4, 1.25) node[above] {\tiny $28$};
  \draw (8, 1.25) node[above] {\tiny $24$};
  \draw (12,1.25) node[above] {\tiny $20$};
  \draw (16,1.25) node[above] {\tiny $16$};
  \draw (20,1.25) node[above] {\tiny $12$};
  \draw (24,1.25) node[above] {\tiny $8$};
  \draw (28,1.25) node[above] {\tiny $4$};
  \draw (32,1.25) node[above] {\tiny $0$};

  \node at (-2,0.5) {\textbf{In}};

  \node[draw, minimum width=2cm, minimum height=0.75cm,xshift=1cm] (in_src3) at (0, 0.5) {\small SRC[3]};
  \node[draw, minimum width=2cm, minimum height=0.75cm,xshift=1cm] (in_src2) at (4, 0.5) {\small SRC[2]};
  \node[draw, minimum width=2cm, minimum height=0.75cm,xshift=1cm] (in_src1) at (8, 0.5) {\small SRC[1]};
  \node[draw, minimum width=2cm, minimum height=0.75cm,xshift=1cm] (in_src0) at (12,0.5) {\small SRC[0]};
  \node[draw, minimum width=2cm, minimum height=0.75cm,xshift=1cm] (in_dst3) at (16,0.5) {\small DST[3]};
  \node[draw, minimum width=2cm, minimum height=0.75cm,xshift=1cm] (in_dst2) at (20,0.5) {\small DST[2]};
  \node[draw, minimum width=2cm, minimum height=0.75cm,xshift=1cm] (in_dst1) at (24,0.5) {\small DST[1]};
  \node[draw, minimum width=2cm, minimum height=0.75cm,xshift=1cm] (in_dst0) at (28,0.5) {\small DST[0]};
  
  \draw (0, -2.25) node[above] {\tiny $32$};
  \draw (4, -2.25) node[above] {\tiny $28$};
  \draw (8, -2.25) node[above] {\tiny $24$};
  \draw (12,-2.25) node[above] {\tiny $20$};
  \draw (16,-2.25) node[above] {\tiny $16$};
  \draw (20,-2.25) node[above] {\tiny $12$};
  \draw (24,-2.25) node[above] {\tiny $8$};
  \draw (28,-2.25) node[above] {\tiny $4$};
  \draw (32,-2.25) node[above] {\tiny $0$};

  \node at (-2,-3) {\textbf{Out}};

  \node[draw, minimum width=2cm, minimum height=0.75cm,xshift=1cm] (out_src3) at (0, -3) {\small SRC[2]};
  \node[draw, minimum width=2cm, minimum height=0.75cm,xshift=1cm] (out_src2) at (4, -3) {\small SRC[1]};
  \node[draw, minimum width=2cm, minimum height=0.75cm,xshift=1cm] (out_src1) at (8, -3) {\small SRC[0]};
  \node[draw, minimum width=2cm, minimum height=0.75cm,xshift=1cm] (out_src0) at (12,-3) {\small $P$};
  \node[draw, minimum width=2cm, minimum height=0.75cm,xshift=1cm] (out_dst3) at (16,-3) {\small 0};
  \node[draw, minimum width=2cm, minimum height=0.75cm,xshift=1cm] (out_dst2) at (20,-3) {\small DST[3]};
  \node[draw, minimum width=2cm, minimum height=0.75cm,xshift=1cm] (out_dst1) at (24,-3) {\small DST[2]};
  \node[draw, minimum width=2cm, minimum height=0.75cm,xshift=1cm] (out_dst0) at (28,-3) {\small DST[1]};

  \draw[-stealth] (in_dst1.south) -- (out_dst0.north);
  \draw[-stealth] (in_dst2.south) -- (out_dst1.north);
  \draw[-stealth] (in_dst3.south) -- (out_dst2.north);

  \draw[-stealth] (in_src0.south) -- (out_src1.north);
  \draw[-stealth] (in_src1.south) -- (out_src2.north);
  \draw[-stealth] (in_src2.south) -- (out_src3.north);

\end{tikzpicture}

%% file: figs/measurement_setup.tex
\begin{tikzpicture}

    \node[draw, minimum width=2cm, minimum height=1cm] (host)       at (-5,0) {Host};
    \node[draw, minimum width=3cm, minimum height=1cm] (switch)     at (0,0)  {Switch};
    \node[draw, minimum width=2cm, minimum height=1cm] (endpoint)   at (5,0)  {Endpoint};

    %switch ports
    \foreach \x in {-1.4,-1,...,1.4}
        \draw[fill] ($(\x, -0.5) + (-0.1,0)$) rectangle ($(\x, -0.7) + (0.1,0)$);
    \draw[fill] (switch.north) -- ++(0.1,0) -- ++(-0.1,0.2) coordinate (switch_trig) -- ++(-0.1,-0.2) -- cycle;

    %host port
    \draw[fill] (-4.5,-0.5) rectangle (-4.3,-0.7);

    %endpoint port
    \draw[fill] (4.5,-0.5) rectangle (4.3,-0.7);
    \draw[fill] (endpoint.north) -- ++(0.1,0) -- ++(-0.1,0.2) coordinate (endpoint_trig) -- ++(-0.1,-0.2) -- cycle;

    %fiber connections
    \draw[thick, red] (-4.35,-0.7) -- ++(0,-0.3) -| node[above, pos=0.25] {Tx $\rightarrow$} (-1.45,-0.7);
    \draw[thick, blue!50]  (-4.45,-0.7) -- ++(0,-0.4) -| node[below, pos=0.25] {Rx $\leftarrow$}  (-1.35,-0.7);

    \draw[thick, red] (1.45,-0.7) -- ++(0,-0.3) -| node[above, pos=0.25] {Tx $\rightarrow$} (4.35,-0.7);
    \draw[thick, blue!50]  (1.35,-0.7) -- ++(0,-0.4) -| node[below, pos=0.25] {Rx $\leftarrow$}  (4.45,-0.7);

    %oscillosope + tappings
    \node[draw, minimum width=4cm, minimum height=2cm] at (0,3) {Oscilloscope};

    \draw[green,stealth-] (1,2)  -- ++(0,-0.5) -| (endpoint_trig);
    \draw[green,stealth-] (0,2)  --               (switch_trig);
    \draw[green,stealth-] (-1,2) -- ++(0,-0.5) -| (-3.8,-1);

    %ethernet
    \draw[<-] (host) -- node[sloped,above]{Ethernet} node[sloped,below]{Trigger} ++(0,3);

\end{tikzpicture}